\documentclass{l4dc2024}
\newtheorem{assumption}{Assumption}
\newtheorem{problem}{Problem}
\newtheorem{fact}{Fact}
\usepackage{times, enumitem}

\title[Bifurcation detection]{Data-Driven Bifurcation Analysis via Learning of Homeomorphism}
\author{%
 \Name{Wentao Tang} \Email{wentao\_tang@ncsu.edu} \\
 \addr Department of Chemical and Biomolecular Engineering, North Carolina State University, \\ Raleigh, NC 27695-7905, USA
}

\begin{document}

\maketitle

\begin{abstract}
    This work proposes a data-driven approach for bifurcation analysis in nonlinear systems when the governing differential equations are not available. 
    Specifically, \emph{regularized regression with barrier terms} is used to learn a \emph{homeomorphism} that transforms the underlying system to a reference linear dynamics -- either an explicit reference model with desired qualitative behavior, or Koopman eigenfunctions that are identified from some system data under a reference parameter value. 
    When such a homeomorphism fails to be constructed with low error, bifurcation phenomenon is detected. 
    A case study is performed on a planar numerical example where a pitchfork bifurcation exists. 
\end{abstract}


\section{Introduction}
\par 
Bifurcation, namely the dependence of qualitative properties on parameters and the loss of structural stability at critical parameter values, is a characteristic behavior of nonlinear dynamical systems \citep{strogatz2015nonlinear}, which often has important engineering consequences (e.g., runaway reactors) \citep{kevrekidis1987numerical, balakotaiah1999bifurcation}. 
This work focuses on \emph{local bifurcations} near an equilibrium point, i.e., the change of local convergence behavior, instead of global/non-local features such as homoclinic or heteroclinic orbits and limit cycles. 
If the system model is available as ordinary differential equations (ODEs), then by constructing a homeomorphism (bijective and continuous mapping with continuous inverse) between the system model and a normal form, bifurcations can be classified into, e.g., fold, Andronov-Hopf, cusp, and Bogdanov-Takens types \citep{guckenheimer2013nonlinear, kuznetsov2023elements}. 
Numerical algorithms can be used for the continuation of ODE solutions to generate accurate bifurcation maps \citep{kuznetsov2019numerical}. 

However, an accurate first-principles model can be difficult to obtain in systems with detailed underlying mechanisms or too complex to be amenable to analysis. In such cases, a \emph{data-driven bifurcation analysis} approach is needed to discover the occurrence of bifurcation when the parameter value varies, thus assisting the system design and control. 
In this work, we assume that accurate state measurements can be obtained (from noiseless simulations), and propose a \emph{regression} approach to detect bifurcation under the following two use cases. 
\begin{itemize}[nolistsep]
    \item In the first case, a reference model is supposed to be available a priori. Such a model has the desired local qualitative behavior of the system and is known to be topologically equivalent (conjugate) to the underlying system under a reference parameter value ($\alpha_0$). Thus, by determining a homeomorphism that transforms the data to conform to this reference model, conclusion can be made on whether under the system bifurcates between the reference and the given parameter value ($\alpha_1$). 
    \item In the second case, a reference ODE model is unavailable. Instead, it is assumed that two datasets are available under $\alpha_0$ and $\alpha_1$. Without the model, empirical Koopman eigenfunctions are assumed to have been identified under $\alpha_0$, which can be achieved via extended mode decomposition \citep{williams2015data}, deep learning \citep{lusch2018deep}, or sparse regression \citep{kaiser2021data}. Hence, a homeomorphism is sought to transform the data under $\alpha_1$ to conform to the linear dynamics of these Koopman eigenfunctions. 
\end{itemize}
The homeomorphism learning problem is formulated as a convex optimization one, with regularization on the coefficient magnitude in the regression problem, as well as barrier terms to enforce the non-degeneracy of the learned homeomorphic mapping. 
When the resulting loss (error) metric is not sufficiently low, local bifurcation is detected. A threshold value for such a bifurcation detection is theoretically obtained in this paper; however, practically, the bifurcation should be considered to occur when the error starts to significantly increase. 

\par The proposed work is related to the recent advances in using sparse regression for finding governing equations \citep{brunton2016discovering}, deep learning for solving dynamical systems \citep{karniadakis2021physics}, as well as on Koopman operators \citep{mauroy2020koopman, otto2021koopman, colbrook2024rigorous} and their applications in control \citep{korda2018linear, huang2022convex, narasingam2023data}. 
We refer to \cite{mezic2020spectrum} for a profound investigation of the spectrum structure of Koopman operators, where it is noted that point spectrum information (eigenfunctions) are often sufficient\footnote{Specifically, it was shown that if a global stabilizing equilibrium exists, then Koopman eigenfunctions can be extended from a small local neighborhood of the equilibrium to the entire attraction basin, and that for limit cycle systems, any function that is $L^2$ in the cyclic coordinate and analytical in the remaining coordinates can be expanded as a series of countably many Koopman eigenfunctions.}. 
This paper is in particular motivated by the work of \cite{bollt2018matching} that focused on ``matching'' dynamics, namely the determination of a homeomorphism between two conjugate systems based on their Koopman eigenfunctions, as well as its extension in \cite{redman2022algorithmic} to determine the equivalence of numerical algorithms as dynamical systems.

\section{Preliminaries and Problem Settings}
Here we consider a parameterized continuous-time dynamical system in the form of
\begin{equation}\label{eq:system}
    \dot{x}(t) = f\left(x(t); \alpha\right) =: f_\alpha(x(t)), 
\end{equation}
where $x(t)\in \mathcal{X}\subseteq \mathbb{R}^n$ is the state vector and $\alpha\in \mathcal{A} \subseteq \mathbb{R}^q$ represents the parameters. We assume that $f$ is smooth to guarantee the existence and uniqueness of solution under any parameter value. Denote the flow of \eqref{eq:system} by a single-parameter group $S_{f_\alpha}^t$, i.e., $x(t_0 + t) = S_{f_\alpha}^t (x(t_0))$. To study the bifurcation behavior, we consider the conjugacy between nonlinear systems.
\begin{definition}
    Two dynamical systems $\dot{x} = f(x)$ and $\dot{y} = g(y)$, on $\mathcal{X}\subseteq \mathbb{R}^n$ and $\mathcal{Y}\subseteq \mathbb{R}^n$, respectively, are said to be \emph{conjugate} if there exists a homeomorphism $h: \mathcal{X}\rightarrow\mathcal{Y}$, such that the flows of these two systems, denoted as $S_f^t$ and $S_g^t$, respectively, satisfy
    \begin{equation}
        S_g^t \circ h = h \circ S_f^t. 
    \end{equation}
\end{definition}
In other words, two conjugate systems are identical via an invertible change of coordinates. Conjugacy is an equivalent relation between $n$-dimensional dynamical systems, i.e., if  $\dot{x} = f(x)$ is known to be conjugate to $\dot{y} = g(y)$, which is further conjugate to another $\dot{z} = k(z)$, then the first system must be conjugate to the last. 

\begin{definition}
    The parametric system \eqref{eq:system} is said to \emph{bifurcate} between parameter values $\alpha_0$ and $\alpha_1\in \mathcal{A}$ if the system at $\alpha_1$ is not conjugate to the system at $\alpha_0$. 
\end{definition}
If the system \eqref{eq:system} at a ``reference parameter'' $\alpha_0$ is conjugate to an a priori ``standard system'' $\dot{y} = g(y)$, then to judge whether the system at $\alpha_1$ is conjugate to the system at $\alpha_0$, it suffices to construct a homeomorphism from the system at $\alpha_1$ and the ``standard system''. 
In this work, we focus on the detection of local bifurcation. Without loss of generality, assume that the equilibrium around which we are interested in is the origin (namely $f_{\alpha}(x) = 0$ for all $\alpha\in \mathcal{A}$). Under mild conditions, the standard system $\dot{y} = g(y)$ can be chosen as the linearization of $\dot{x} = f_{\alpha_0}(x)$ at $x=0$. 
\begin{assumption}\label{assum:1}
    Let $A = \partial f_{\alpha_0}(0)/\partial x$ (the Jacobian matrix). We assume that the eigenvalues of $A$, $(\lambda_1, \dots, \lambda_n)$, are of type $(C, \nu)$ for some $C, \nu > 0$. That is, for any $\lambda_k$ and $(m_1, \dots, m_n)\in \mathbb{N}^n$ with $\sum_{r=1}^n m_r \geq 2$, the following inequality holds: $ |\lambda_k - \sum_{r=1}^n m_r\lambda_r| \geq C (\sum_{r=1}^n m_r)^{-\nu}$.
\end{assumption}
\begin{fact}[cf. \cite{arnold1988geometrical}, \S24]
    Under the above assumption, the system \eqref{eq:system} at $\alpha_0$ is conjugate to the linearized system $\dot{y} = Ay$. 
\end{fact}

Now our first problem in consideration is stated as follows. 
\begin{problem}\label{prob:1}
    Assume that the system $\dot{y} = g(y) = Ay$ on $\mathbb{R}^n$ is given. Suppose that at $\alpha = \alpha_1$, a dataset comprising of (sufficiently many) orbits, sampled at a frequency of $1/\tau$:
    \begin{equation}\label{eq:dataset}
        \mathcal{D} = \left\{ \left(x^{(i)}_0, x^{(i)}_1, \dots, x^{(i)}_P \right) \Big{|} i=1, \dots, M\right\}
    \end{equation}
    is available, where $x^{(i)}_k = S_{f_\alpha}^\tau(x^{(i)}_{k-1})$ for $i=1,\dots,M$ and $k=1,\dots,P$. Learn a homeomorphism $h: \mathcal{X}\rightarrow \mathcal{Y}$ such that the following equations approximately hold:
    \begin{equation}\label{eq:matching.1}
    h\left(x^{(i)}_k\right) = e^{A\tau} h\left(x^{(i)}_{k-1}\right), \quad i=1,\dots,M, \enskip k=1,\dots,P. 
    \end{equation}
\end{problem}

\par However, it is not always realistic to assume that the model at $\alpha_0$ is known so that the Jacobian $A$ can be easily obtained. We consider a second setting where $n$ eigenfunctions of the Koopman operators are known, or can be first identified from data, for the system at $\alpha_0$. 
\begin{definition}
    The \emph{Koopman operators}, denoted as $U_{f_\alpha}^t$, for the parametric system \eqref{eq:system}, refer to the family of linear operators on the space of continuous functions on $\mathcal{X}$, such that
    \begin{equation}
        U_{f_\alpha}^t \circ \psi = \psi \circ S_{f_\alpha}^t, \quad \forall \psi \in \mathcal{C}(\mathcal{X}). 
    \end{equation}
    Here any $\psi$ is called an \emph{observable}. We say that $\psi$ is an \emph{eigenfunction} of the Koopman operators associated with (complex-valued) eigenvalue $\lambda$, if 
    \begin{equation}
        (U_{f_\alpha}^t \circ \psi)(x) = e^{\lambda t}\psi(x), \quad \forall x \in \mathcal{X},  
    \end{equation}
    and a \emph{generalized eigenfunction}, if for some integer $k \geq 1$, $\left(d/dt - \lambda\right)^k \psi(x) = 0$. 
\end{definition}

In other words, the eigenfunction $\psi$ is an observable whose value gains a factor of $e^{\lambda \tau}$ as time proceeds by $\tau$. It is natural that if two dynamical systems $\dot{x} = f(x)$ and $\dot{y} = g(y)$ are conjugate through a homeomorphism $h$ (i.e., $y = h(x)$), then from any eigenfunction $\psi$ of the former system, $\psi\circ h^{-1}$ specifies an eigenfunction of the latter system associated with the same eigenvalue. 
For the parametric system \eqref{eq:system} at $\alpha_0$, if it is conjugate to a linear system $\dot{y} = Ay$, then the dynamics can be fully described by the eigenfunctions of the latter. For this linear system, the eigenfunctions are related to the Jordan canonical form of the matrix $A$.  
\begin{fact}[cf. \cite{mezic2020spectrum}, \S 3]
    If a set of functions $\psi = (\psi_1, \dots, \psi_n)$ satisfy $\dot{\psi}(x) = J\psi(x)$, where $J$ is the Jordan canonical form of $A$, then they are generalized Koopman eigenfunctions of the linear system $\dot{y} = Ay$. Moreover, these generalized eigenfunctions come from an invertible transform of $y$; specifically $\psi_i = w_i^\top y$ ($i=1,\dots,n$), with $\{w_1, \dots, w_n\}$ being a basis of $\mathbb{R}^n$. 
\end{fact}

Without knowing the model at $\alpha_0$, the eigenfunctions can be identified from data instead. 
A common approach to this end is extended dynamic mode decomposition (EDMD) \citep{williams2015data}, where a dictionary of basis functions on $x$ is used to lift the states $x$ into a high-dimensional space, and then a linear dynamics $A$ is found via regression. 
Alternatively, instead of obtaining $A$, a sparse regression on dictionary functions can be performed under any given $\lambda$ to determine a Koopman eigenfunction \citep{kaiser2021data}. 
It has been proved that for the Koopman operator estimated on a finite number $N$ of basis functions and sample size $M$, as $M\rightarrow\infty$, it will converge to the optimal approximation, \citep{klus2016numerical}, and as $N\rightarrow \infty$, the finite-rank approximation will converge to the true Koopman operator in strong operator topology \citep{korda2018convergence}. 

\par Hence, finding the Koopman eigenfunctions at $\alpha = \alpha_1$ that evolve according to the same linear dynamics as the data-identified eigenfunctions at $\alpha = \alpha_0$ serves as an approximation for detecting bifurcation. 
The problem we are interested in is follows. 
\begin{problem}\label{prob:2}
    Assume that the eigenfunctions $\psi = (\psi_1, \dots, \psi_n)$ are known to satisfy $\dot{\psi} = A\psi$ for a given $A$ for the system \eqref{eq:system} at $\alpha = \alpha_0$. Suppose that at $\alpha_1$, a (sufficiently large) dataset in the form of \eqref{eq:dataset} is available. Learn a homeomorphism $h: \mathcal{X}\rightarrow \mathcal{Y}$ such that \eqref{eq:matching.1} holds. 
\end{problem}

\par Based on above discussions, detecting local bifurcation around a given equilibrium based on a known reference model $\dot{x} = f_{\alpha_0}(x)$ and based on Koopman eigenfunctions $(\psi_1, \dots, \psi_n)$ identified from data can have the same formulation. With Assumption \ref{assum:1}, which allows the reference system to be conjugate to the linearized system, the only difference between the two problems is the assignment of matrix $A$. 
Now, focusing on \eqref{eq:matching.1}, we aim at determining the homeomorphism $h$ from data through a regression approach for both the \emph{model-aware} Problem \ref{prob:1} and \emph{model-free} Problem \ref{prob:2}.

\section{Proposed Approach}
\subsection{Least Squares Formulation}
\par Now we choose a basis $\{\varphi_1, \dots, \varphi_N\}$ for the components of $h$, i.e., to approximate $h$ by
\begin{equation}
    \begin{bmatrix}
        h_1(x) & \cdots & h_n(x) 
    \end{bmatrix} = 
    \begin{bmatrix}
        \varphi_1(x) & \cdots & \varphi_N(x) 
    \end{bmatrix}
    \begin{bmatrix}
        \theta_{11} & \cdots & \theta_{1n}(x) \\
        \vdots & \ddots & \vdots \\
        \theta_{N1} & \cdots & \theta_{Nn}(x) \\
    \end{bmatrix}, 
\end{equation}
namely $h_j(x) = \sum_{k=1}^N \theta_{kj}\varphi_k(x)$ and $h(x) = h_\Theta(x) := \Theta^\top \varphi(x)$, where $\Theta\in \mathbb{R}^{N\times n}$ is the coefficient matrix to be learned and $\varphi = [\varphi_1, \dots, \varphi_N]^\top$. 
Choosing a polynomial basis as $\varphi$ is desirable, due to the approximability of any continuous function by polynomials in $\mathcal{C}^\infty$ (Stone-Weierstrass theorem) as well as in Sobolev space (Meyers-Serrin theorem). The latter property is needed since we need $h$ to be an invertible mapping, which boils down to constraining the Jacobian of $h$ to be always non-singular (assuming that $h$ is differentiable). 
With this parameterization, the error on the $i$-th sampled orbit $(x^{(i)}_0, \dots, x^{(i)}_P)$ resulted from the parameters $\Theta$ is thus the collection of vectors
\begin{equation}
    e^{(i)}_k(\Theta) = \Theta^\top \varphi\left( x^{(i)}_k \right) - e^{A\tau} \Theta^\top \varphi\left( x^{(i)}_{k-1} \right), \quad i=1,\dots,M, \enskip k=1,\dots,P. 
\end{equation}
Denoting 
\begin{equation}
    \Phi_k = \begin{bmatrix}
        \vdots \\ \varphi\left( x^{(i)}_k \right)^\top \\ \vdots
    \end{bmatrix} \in \mathbb{R}^{M\times N}, \enskip
    E_k(\Theta) = \begin{bmatrix}
        \vdots \\ e^{(i)}_k(\Theta)^\top \\ \vdots
    \end{bmatrix} \in \mathbb{R}^{M\times n}, 
\end{equation}
and $\exp(A\tau) = \bar{A}$, then we have $E_k(\Theta) = \Phi_k\Theta - \Phi_{k-1}\Theta \bar{A}^\top$ for $k=1,\dots,P$. Hence, we can have a least square formulation of minimizing the total squared Frobenius norm of $E_1, \dots, E_P$, scaled by the length of prediction horizon $P$ and sample size $M$:
\begin{equation}
    \min_{\Theta} \enskip 
    \frac{1}{2MP}\sum_{k=1}^P \|E_k(\Theta)\|_\mathrm{F}^2 = 
    \frac{1}{2MP}\sum_{k=1}^P \| \Phi_k\Theta - \Phi_{k-1}\Theta \bar{A}^\top \|_\mathrm{F}^2. 
\end{equation}

\subsection{Regularization and Barrier Terms}
\par Here we use a Tikhonov regularization on the coefficients, with $\beta$ being the regularization hyperparameter, in order to smooth the contours of $h$ to mitigate generalization loss. (The $\ell_2$-regularization can be replaced by a LASSO term, which seems to be unnecessary as sparsity of non-zero coefficients in $h$ does not have a particular meaning.) 
To enforce the invertibility of $h_\Theta := \Theta^\top \varphi$, we assume that all basis functions are differentiable (which is obviously true for polynomials) and require that the Jacobian matrix of $h_\Theta$ evaluated at each $x^{(i)}_0$ must have full rank. 
That is, by letting $C^{(i)} = \partial \varphi(x^{(i)}_0)/\partial x$, we require that $\Theta^\top C^{(i)}$ must not have zero singular values for all $i=1,\dots,M$, i.e., $\Theta^\top C^{(i)}C^{(i)\top}\Theta \succ 0$. 
To impose this positive definiteness constraint, we adopt a log-barrier term with barrier hyperparameter $\mu > 0$. This results in additional objective terms: $-\mu \log\det (\Theta^\top C^{(i)}C^{(i)\top}\Theta)$ defined on all sample points $i=1,\dots,M$. 

\par However, the log-barrier term above, despite being convex, can be hardly handled by solvers (e.g., \texttt{cvxpy}) due to the quadratic dependence on $\Theta$ in the log-determinant. For a practical treatment, in view of the relation $\Theta^\top C^{(i)} C^{(i)\top}\Theta \succeq \Theta^\top C^{(i)} + C^{(i)\top}\Theta - I$, we istead define the barrier terms: $-\mu \log\det (\Theta^\top C^{(i)} + C^{(i)\top}\Theta - I)$. 
It should also be noted that such barrier terms have the side effect of rewarding $h$ to have a Jacobian with large determinant. To mitigate this effect, we force penalization term to be nonnegative. 
Now, the optimization problem to be solved is as follows. 
\begin{equation}\label{eq:optimization}
\begin{aligned}
    \min_{\Theta \in \mathbb{R}^{N\times n}} \enskip J_\mu(\Theta) = & \frac{1}{2MP}\sum_{k=1}^P \| \Phi_k\Theta - \Phi_{k-1}\Theta \bar{A}^\top \|_\mathrm{F}^2 + \frac{1}{2}\beta\|\Theta\|_\mathrm{F}^2 \\
    &+ \frac{\mu}{M}\sum_{i=1}^M \max\left\{0, -\log \det\left( \Theta^\top C^{(i)} + C^{(i)\top}\Theta - I \right) \right\}. 
\end{aligned}
\end{equation}

\par 
In this convex optimization formulation \eqref{eq:optimization}, $\mu$ is considered as a tunable hyperparameter that gives a tradeoff between the mean squared error (MSE) plus the Tikhonov regularization cost versus the Jacobian magnitude. To select an optimal value $\mu = \mu_\ast$, we consider a \emph{scaled MSE} after the optimization problem above on $J_\mu(\Theta)$ is solved: 
\begin{equation}\label{eq:MSE}
    L_\mu := \frac{\left[ \frac{1}{MP}\sum_{k=1}^P \| \Phi_k\Theta - \Phi_{k-1}\Theta \bar{A}^\top \|_\mathrm{F}^2 \right]^{1/2}}{\frac{1}{M} \sum_{i=1}^M |\det\left(\Theta^\top C^{(i)}\right)|},
\end{equation}
in which the denominator is interpreted as a stochastic approximation of the measure of the mapped state space. Specifically, suppose that the sample of initial states $\{x^{(i)}_0| i=1,\dots,M\}$ are obtained from a probability distribution $\rho$ on $\mathcal{X}$. Then the denominator is in fact
\begin{equation}
    \frac{1}{M}\sum_{i=1}^M \bigg{|}\frac{\partial h(x^{(i)}_0)}{\partial x}\bigg{|} \approx 
    \int_{\mathcal{X}} \bigg{|} \frac{\partial h(x)}{\partial x}\bigg{|} d\rho(x) = \rho_\ast(\mathcal{Y}),  
\end{equation}
in which $|\cdot|$ for a Jacobian matrix is used to denote the absolute value of its determinant and $\rho_\ast$ is the push-forward of $\rho$ onto $\mathcal{Y}$ under $h$. In particular, if the sampling is uniform on a compact $\mathcal{X}$, then $\rho_\ast(\mathcal{Y}) = \mathrm{mes}(\mathcal{Y}) / \mathrm{mes}(\mathcal{X})$, where $\mathrm{mes}$ is the Lebesgue measure. 
Hence, the optimal barrier hyperparameter $\mu_\ast$ is chosen \emph{a posteriori} according to the minimal $L_\mu$ in \eqref{eq:MSE}. 

\par 
Using the scaled MSE $L_\mu$ as the posterior loss metric, the regularization parameter $\lambda$ is also tunable. Specifically, $K$-fold cross-validation is used to evaluate the effect of regularization. That is, the entire dataset is first partitioned into $K$ portions. Then, in each of the $K$ parallel training experiments, one portion is used for the validation following the optimization on the remaining data. The geometric mean of the $K$ generalization losses is taken as the validated  $L_\mu$ under the specific $\lambda$ value. 
Furthermore, the richness of polynomial basis function, specified by the maximum allowed degree $d$, is also tuned according to the training loss and validation loss. 
The sample size $M$ should be moderately large, but not excessive so as to reduce the computational time for optimization. The same principle applies to the selection of prediction horizon $P$ and sampling frequency $1/\tau$.

\subsection{Generalization Loss}
In this subsection, we aim to establish theoretical probabilistic upper bounds on the generalized MSE and cumulative prediction error, supposing that bifurcation does not occur, i.e., the actual dynamics can be matched to $A$. Now let us denote by $L_\ast$ the posterior minimized $L_\mu$ value at $\mu_\ast$, $\Theta_\ast$ the optimized parameter value under $\mu_\ast$, and $h_\ast = \Theta_\ast^\top \varphi$ the corresponding homeomorphic mapping learned. According to \eqref{eq:MSE}, we have 
\begin{equation}
    \frac{1}{MP} \sum_{k=1}^P \|\Phi_k\Theta - \Phi_{k-1}\Theta A^\top\|_\mathrm{F}^2 \leq L_\ast^2
    \left[ \frac{1}{M}\sum_{i=1}^M \bigg{|}\det \left(\Theta^\top C^{(i)} \right)\bigg{|} \right]^2,  
\end{equation}
which boils down to 
\begin{equation}\label{eq:generalization.1}
    \frac{1}{MP}\sum_{i=1}^M \sum_{k=1}^P \left\| h(x^{(i)}_k) - \bar{A} h(x^{(i)}_{k-1}) \right\|^2 \leq 
    L_\ast^2 \left[ \frac{1}{M}\sum_{i=1}^M \bigg{|}\frac{\partial h(x^{(i)}_0)}{\partial x}\bigg{|} \right]^2. 
\end{equation}

\begin{assumption}
    The initial states $\{x^{(i)}_0\}_{i=1}^M$ are independently sampled from a bounded $\mathcal{X}$. 
\end{assumption}
Since $\mathcal{X}$ is bounded, according to Hoeffding's inequality, with confidence $1-\delta_1$, we have
\begin{equation}
    \mathbb{E} \left[ \frac{1}{P}\sum_{k=1}^P \left\| \left(h_\ast\circ S_{f_\alpha}^{k\tau}\right)(x) - \bar{A} \left(h_\ast\circ S_{f_\alpha}^{(k-1)\tau}\right)(x) \right\|^2 \right] \leq 
    \text{L.H.S. of \eqref{eq:generalization.1}} + F_\ast\sqrt{\frac{\log (1/\delta_1)}{2M}},
\end{equation}
and with confidence $1-\delta_2$, we have
\begin{equation}
    \frac{1}{M}\sum_{i=1}^M \bigg{|}\frac{\partial h_\ast(x^{(i)})}{\partial x}\bigg{|} \leq \mathbb{E}\left[\bigg{|} \frac{\partial h_\ast(x)}{\partial x}\bigg{|} \right] + H_\ast\sqrt{\frac{\log(1/\delta_2)}{2M}}, 
\end{equation}
where the expectation is on the population $x\sim \rho$ and 
$$F_\ast := \sup_{x\in\mathcal{X}} \frac{1}{P}\sum_{k=1}^P \left\| \left(h_\ast\circ S_{f_\alpha}^{k\tau}\right)(x) - \bar{A} \left(h_\ast\circ S_{f_\alpha}^{(k-1)\tau}\right)(x) \right\|^2, \quad
H_\ast := \sup_{x\in\mathcal{X}} \bigg{|} \frac{\partial h_\ast(x)}{\partial x}\bigg{|}. $$
Thus, with confidence $1-\delta$ where $\delta = \delta_1 + \delta_2$, we have for the system at $\alpha$ the following estimation of the generalization error on matching the target linear dynamics $A$: 
\begin{equation}
\begin{aligned}
    & \mathbb{E} \left[ \frac{1}{P}\sum_{k=1}^P \left\| \left(h_\ast\circ S_{f_\alpha}^{k\tau}\right)(x) - \bar{A} \left(h_\ast\circ S_{f_\alpha}^{(k-1)\tau}\right)(x) \right\|^2 \right] \leq \\
    & \qquad\qquad L_\ast^2 \left\{ \mathbb{E}\left[\bigg{|} \frac{\partial h_\ast(x)}{\partial x}\bigg{|} \right] + H_\ast \sqrt{\frac{\log(1/\delta_2)}{2M}} \right\}^2 + F_\ast \sqrt{\frac{\log(1/\delta_1)}{2M}}. 
\end{aligned}
\end{equation}
Therefore, we have the following conclusion. 
\begin{theorem}
    Under the previous assumptions, if the system \eqref{eq:system} does not bifurcate between $\alpha_0$ and $\alpha$, then when $M$ is sufficiently large, with confidence $1-\delta$, 
    \begin{equation}
        \mathbb{E} \left[ \frac{1}{P}\sum_{k=1}^P \left\| \left(h_\ast\circ S_{f_\alpha}^{k\tau} - \bar{A} h_\ast\circ S_{f_\alpha}^{(k-1)\tau}\right)(x) \right\|^2 \right] \leq 
        L_\ast^2 \mathbb{E}\left[\bigg{|} \frac{\partial h_\ast(x)}{\partial x}\bigg{|} \right]^2 
        + \mathcal{O}\left(\sqrt{\frac{\log(1/\delta)}{M}} \right).
    \end{equation}
\end{theorem}
In particular, we consider the case when $A$ is Hurwitz, so that $\|\bar{A}\|\leq \alpha$ for some $\alpha \in (0, 1)$. In this case, the expectation term on the left-hand side in the above theorem can be converted into the \emph{cumulative squared prediction error} starting from the initial state $x$, with the coefficient $C_\alpha$ being the $\ell_2$-gain of the discrete-time linear system $\bar{A}$, which depends on $\alpha$. It is possible to further bound the expected continuous-time cumulative prediction error, which is not detailed here. 
\begin{corollary}\label{corollary}
    If $\|\bar{A}\|\leq \alpha < 1$, then under the previous assumptions, if the system \eqref{eq:system} does not bifurcate between $\alpha_0$ and $\alpha$, then at large $M$, with confidence $1-\delta$, 
    \begin{equation}
        \mathbb{E} \left[ \frac{1}{P}\sum_{k=1}^P \left\| \left(h_\ast\circ S_{f_\alpha}^{k\tau} - \bar{A}^k h_\ast\right) (x) \right\|^2 \right] \leq 
        C_\alpha L_\ast^2 \mathbb{E}\left[\bigg{|} \frac{\partial h_\ast(x)}{\partial x}\bigg{|} \right]^2 + \mathcal{O}\left(\sqrt{\frac{\ln(1/\delta)}{M}} \right). 
    \end{equation}
\end{corollary}
    
\par Clearly, the generalization loss mainly depends on the optimized loss $L_\ast$, which is a scaled MSE. The magnitude of this quantity is therefore of interest. Indeed, if bifurcation does not occur and $h_\ast$ can be obtained from a dense infinite-dimensional subspace of all continuous functions on $\mathcal{X}$, then $L_\ast$ can approach $0$ with a ground-truth solution. 
However, as the components of $h$ are restricted to polynomials of degree not exceeding $d$, there exists an optimal parameter choice $\Theta_\circ$ and a corresponding $h_\circ = h_{\Theta_\circ}$ to achieve the lowest possible error 
\begin{equation}
    \epsilon_d^2 := \min_\Theta \mathbb{E} \enskip \left[ \frac{1}{P}\sum_{k=1}^P \left\| \left(h\circ S_{f_\alpha}^{k\tau} - \bar{A}  h\circ S_{f_\alpha}^{(k-1)\tau}\right)(x) \right\|^2 \right] 
\end{equation}
among all such degree-constrained $h$ that makes $|\partial h_\circ(x)/\partial x| \geq 1$ almost everywhere. 
Hence for this specific feasible solution, $h_\circ$, the barrier term in the objective function \eqref{eq:optimization} is zero and thus independent of $\mu$. Denote this feasible objective as $J_\circ$. Hence, at any given $\mu$, after solving \eqref{eq:optimization} to the optimum, the objective $J_\mu$ and the corresponding solution $\Theta_\mu$ must be such that 
\begin{equation}\label{eq:bound.in.barrier}
    \frac{1}{2MP} \sum_{k=1}^P \|\Phi_k\Theta_\mu - \Phi_{k-1}\Theta_\mu \bar{A}^\top\|_\mathrm{F}^2 + \frac{\mu}{M}\sum_{i=1}^M \max\left\{ 0, -\ln\det\left( \Theta_\mu^\top C^{(i)} + C^{(i)\top} \Theta_\mu - I \right) \right\}
    \leq J_\circ.
\end{equation}
It follows, by relaxing the first summation term, that 
\begin{equation}
    \frac{1}{M}\sum_{i=1}^M \left[1 - |\det ( \Theta_\mu^\top C^{(i)} )|\right] \leq \frac{1}{2\mu} J_\circ
    \quad \overset{\eqref{eq:MSE}}{\Rightarrow} \quad 
    L(\mu) \leq \frac{\sqrt{2J_\circ}}{1 - \frac{1}{2\mu} J_\circ}, 
\end{equation}
and thus, by taking the infimum over $\mu>0$, $L_\ast^2 \leq  2J_\circ$. Similar to the previous discussions, we give a probabilistic upper bound on the right-hand side, $2J_\circ$, which is the MSE reached by $\Theta_\circ$. That is, with confidence $1-\delta$, 
\begin{equation}\label{eq:bound.in.L}
    L_\ast^2 \leq \epsilon_d^2 + F_d\sqrt{\frac{\ln (1/\delta)}{2M}} + \beta \|\Theta_\circ\|_{\mathrm{F}}^2, 
\end{equation}
where 
$$F_d := \sup_{x\in\mathcal{X}} \frac{1}{P}\sum_{k=1}^P \left\| \left(h_\circ \circ S_{f_\alpha}^{k\tau}  - \bar{A} h_\circ \circ S_{f_\alpha}^{(k-1)\tau}\right)(x) \right\|^2.$$

\begin{theorem}
    By solving the barrier optimization problem \eqref{eq:optimization} and tuning the barrier constant $\mu$ according to the criterion \eqref{eq:MSE}, the learned mapping $h$ has the following generalization loss on the linear dynamics with confidence $1-\delta$:
    \begin{equation}
    \begin{aligned}
        & \mathbb{E} \left[ \frac{1}{P}\sum_{k=1}^P \left\| \left(h_\ast\circ S_{f_\alpha}^{k\tau} - \bar{A} h_\ast\circ S_{f_\alpha}^{(k-1)\tau}\right)(x) \right\|^2 \right] \leq \\
        & \qquad\qquad \left[\epsilon_d^2 + F_d\sqrt{\frac{\ln (1/\delta)}{2M}} + \beta \|\Theta_\circ\|_{\mathrm{F}}^2\right] \mathbb{E}\left[\bigg{|} \frac{\partial h_\ast(x)}{\partial x}\bigg{|} \right] + \mathcal{O}\left(\sqrt{\frac{\ln(1/\delta)}{M}} \right). 
    \end{aligned}
    \end{equation}
\end{theorem}
In the sample size limit ($M\rightarrow \infty$), the linearization error is caused by $\epsilon_d$ and $F_d$ (which should converge to $0$ as $d\rightarrow\infty$) as well as $\beta$, the regularization parameter. Similar to the Corollary \ref{corollary}, if $A$ is stable, then the left-hand side above can be replaced by the expected cumulative prediction error.

\section{Numerical Case Study}
\par Now we consider the following planar system example (Example 8.1.3, \cite{strogatz2015nonlinear}):
\begin{equation}
    \dot{x}_1 = \alpha x_1 + x_2 + (1/2\pi)\sin (2\pi x_1), \quad \dot{x}_2 = x_1 - x_2. 
\end{equation}
When $\alpha < -2$, the origin is a stable equilibrium; when $\alpha > -2$, the origin becomes a saddle and two other stable equilibria appear. This is known as a supercritical pitchfork bifurcation. At $\alpha_0 = -4$, the linearized system at the origin gives two negative eigenvalues: $-2 \pm \sqrt{3}$. We will examine the existence of local bifurcation as $\alpha_1$ sweeps from $-4$ to $-1$. The phase diagram for $\alpha = -4$, $-2$, and $-1$ are shown in Fig. \ref{fig:phase}. In the model-aware setting, bifurcation detection aims at constructing $h$ to match the linearized dynamics with $A = \mathrm{diag}(-2 - \sqrt{3}, -2 + \sqrt{3})$.

\vspace{-1em}
\begin{figure}[ht]
\floatconts{fig:phase}{\vspace{-2em}\caption{Phase diagram at different $\alpha$ values.}}{%
\subfigure[$\alpha=-4$][b]{
    \label{fig:phase.4}
    \includegraphics[width=1.85in]{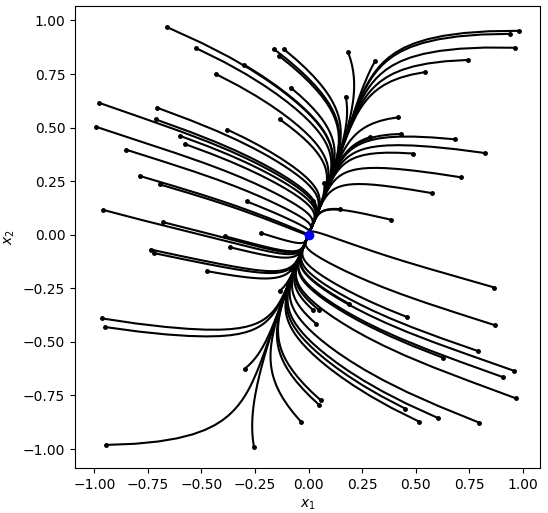}
} 
\subfigure[$\alpha=-2$][b]{
    \label{fig:phase.2}
    \includegraphics[width=1.85in]{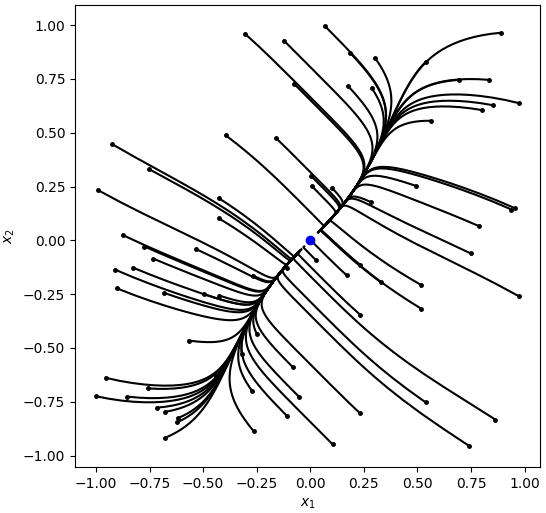}
} 
\subfigure[$\alpha=-1$][b]{
    \label{fig:phase.1}
    \includegraphics[width=1.85in]{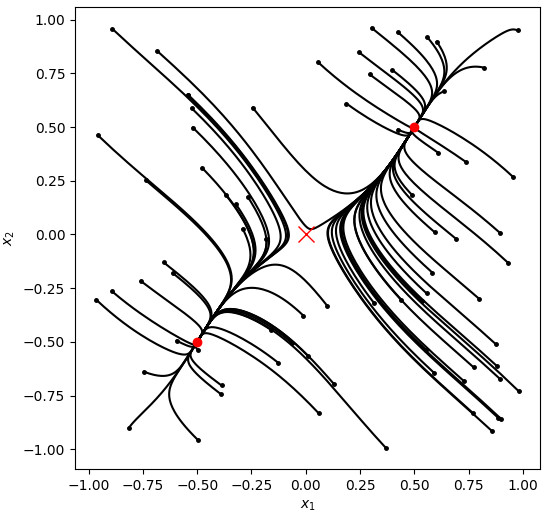}
}
}
\end{figure}

\vspace{-1em}
\par
\begin{figure}[ht]
\floatconts{fig:Koopman}{\vspace{-2em}\caption{Koopman eigenfunctions constructed from data by varying eigenvalue.}}{%
\subfigure[$\mathrm{RMSE}$ versus $\lambda$][b]{
    \label{fig:Koopman.training}
    \includegraphics[width=1.8in]{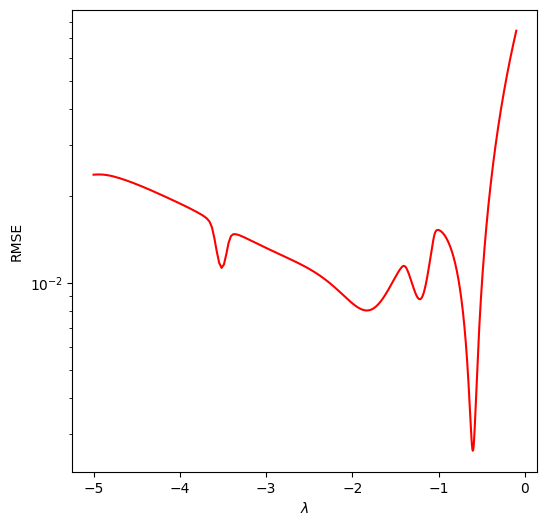}
} 
\subfigure[$\psi_1(x)$ for $\lambda_1=-3.4995$][b]{
    \label{fig:Koopman.1}
    \includegraphics[width=1.9in]{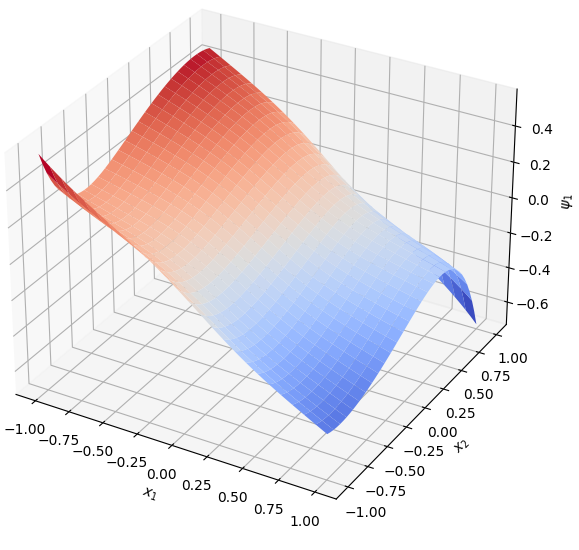}
} 
\subfigure[$\psi_2(x)$ for $\lambda_2=-0.5970$][b]{
    \label{fig:Koopman.2}
    \includegraphics[width=1.9in]{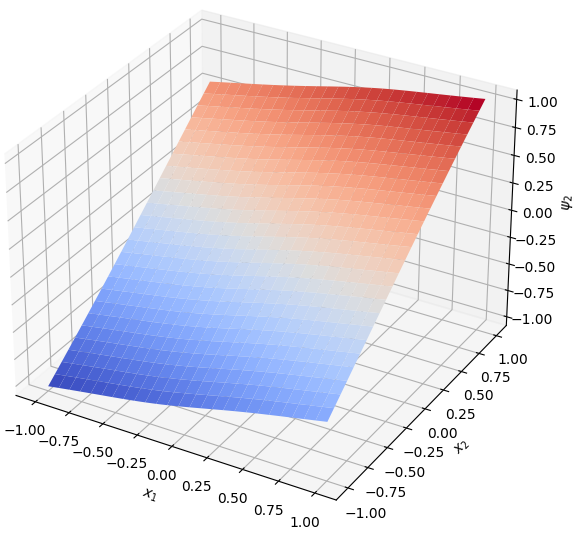}
}
}
\end{figure}

 Assuming that the model is unavailable, through a data-driven approach \citep{kaiser2021data}, we may seek  eigenfunctions at any given eigenvalue $\lambda$ (using polynomials of degree up to $5$). Such eigenvalues should be chosen at the values where the linear dynamics show a minimal RMSE (root mean square error). From Fig. \ref{fig:Koopman.training}, where $\lambda$ ranges from $-5$ to $-0.1$, it is seen that the RMSE has four local minima: at $\lambda_1 = -3.4995$, $\lambda_2 = -0.5970$, and two other points in between. We choose the eigenvalues at the two ends that best avoid dependence between $\psi_1$ and $\psi_2$, as illustrated in Figs. \ref{fig:Koopman.1} and \ref{fig:Koopman.2}. The convergence of $(\psi_1, \psi_2)$ to zero explains the attraction of states to the origin in two dimensions as shown in Fig. \ref{fig:phase.4}. In the model-free setting, bifurcation detection needs to match the Koopman-linearized dynamics with $A = \mathrm{diag}(-3.4995, -0.5970)$.

\vspace{-1em}

\par The barrier constant $\mu$, regularization constant $\beta$, and prediction horizon $P$ for bifurcation detection are tuned according to the MSE term (red circles), regularization term (red triangles) and cross-validated $L_\ast^2$ value (blue squares), as shown in Fig. \ref{fig:tuning}. 
The tuning is performed under $\alpha = \alpha_0 = -4$. The following hyperparameters are chosen: $\mu = 10^{-3}$ (when $L_\ast$ has almost settled on a minimal value) and $\beta = 10^{-4}$ (when the regularization term is slightly less than the MSE term). 
For the choice of $P$, since $\tau = 1/10$, it is noted that when $P = 12$ (i.e., with a prediction horizon of $1.2$ time units), the MSE term and regularization term become close. When $P>20$, the effect of $P$ becomes small -- while the accuracy is improved, longer computational time is needed for high $P$. Hence, $P=5$, $10$, $15$, $20$, and $25$ will be used for bifurcation detection. 

\begin{figure}[ht]
\floatconts{fig:tuning}{\vspace{-2em}\caption{Hyperparameter tuning.}}{%
\subfigure[Barrier constant][b]{
    \label{fig:tuning.barrier}
    \includegraphics[width=1.9in]{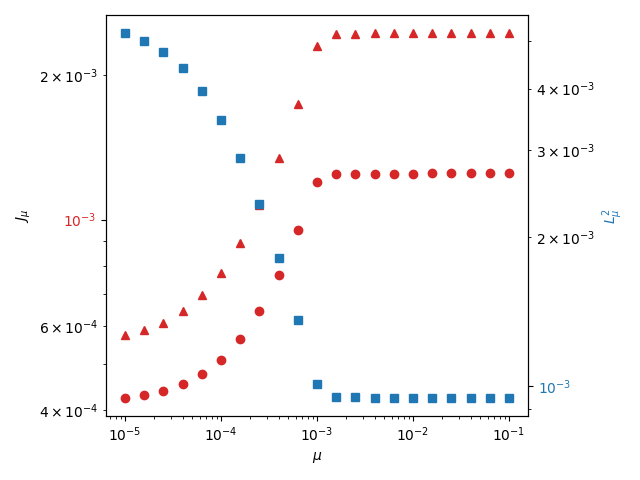}
} 
\subfigure[Regularization constant][b]{
    \label{fig:tuning.regularization}
    \includegraphics[width=1.9in]{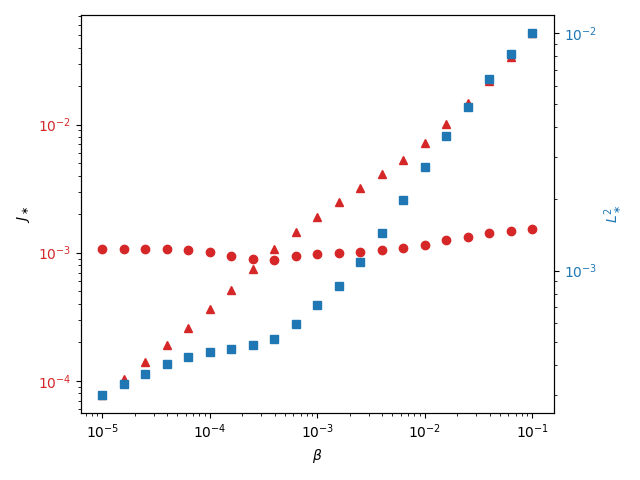}
} 
\subfigure[Prediction horizon][b]{
    \label{fig:tuning.prediction}
    \includegraphics[width=1.9in]{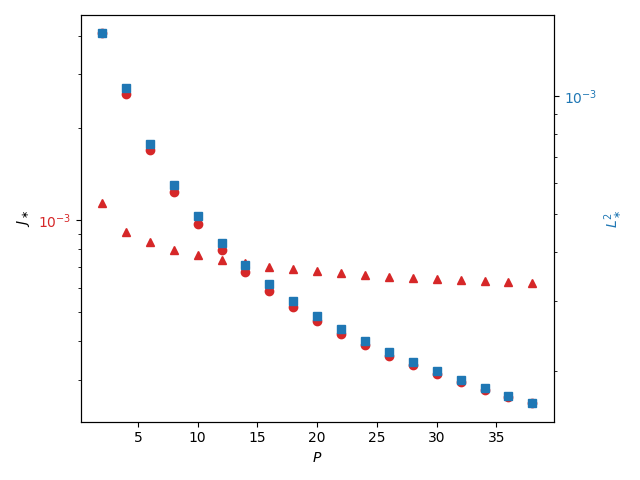}
}
}
\end{figure}

\par
\begin{figure}[ht]
\floatconts{fig:bifurcation}{\vspace{-2em}\caption{Bifurcation detection as the system parameter $\alpha$ varies.}}{%
\subfigure[Model-based $A$][b]{
    \label{fig:bifurcation.modelbased}
    \includegraphics[width=2.0in]{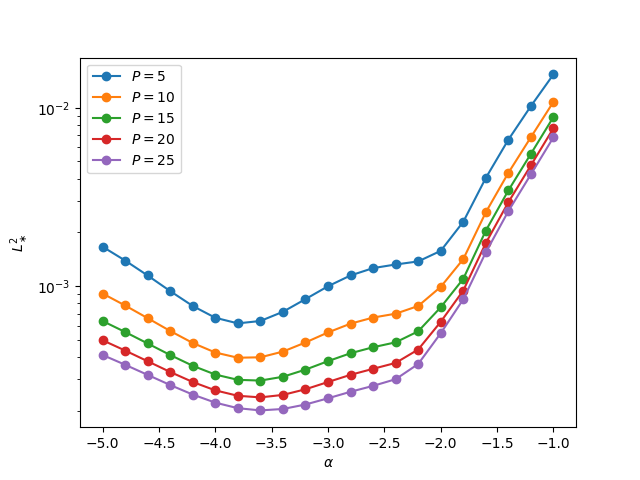}
} 
\subfigure[Koopman-based $A$][b]{
    \label{fig:bifurcation.modelfree}
    \includegraphics[width=2.0in]{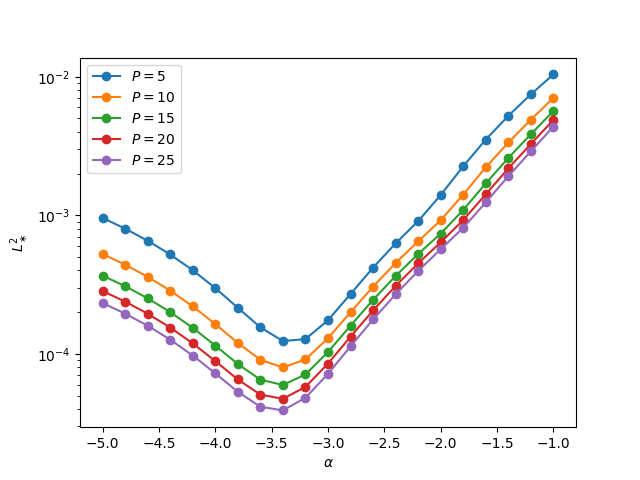}
}
}
\end{figure}

 Using the proposed method, we sample $M=100$ initial points in $\mathcal{X} = [-1, 1]\times [-1, 1]$ and perform the homeomorphism learning under varying values of $\alpha_1$ from $-5$ to $-1$ with an increment of $0.2$. Both model-aware and model-free (Koopman eigenfunction-based) settings are considered. The dependence of loss $L_\ast^2$ on system parameter $\alpha$ is shown in Fig. \ref{fig:bifurcation.modelbased} and \ref{fig:bifurcation.modelfree}, respectively, for these two cases. When $\alpha > -2$, $L_\ast^2$ increases exponentially with $\alpha$, reflecting the occurrence of bifurcation. 
However, in the model-free setting, such an exponential growth is triggered before the bifurcation actually takes place, and the value of $L_\ast^2$ is sensitive to $\alpha$ even without bifurcation. In contrast, in the model-based setting, the behavior before and after bifurcation are highly different, demonstrating that for local bifurcation detection, if the local linearization $A = \partial f_{\alpha_0}(0) / \partial x$ is available, this $A$ matrix is a better linear dynamics to be matched. 
On the other hand, the lower value of $L_\ast^2$ in the model-free case explains that when using Koopman eigenfunctions, one can obtain better linearization in the provided region $\mathcal{X}$ under the reference parameter value $\alpha_0=-4$. But such a Koopman linearization appears to be less robust to parameter variations. 

\vspace{-1em}
\vspace{-1em}

\section{Conclusion}
In this paper, a data-driven method of determining local bifurcation is proposed, where it is assumed that a linear model is available at the reference system parameter value, either by taking the Jacobian of the model or by identifying the Koopman eigenfunctions. 
Specifically, in the absence of bifurcation, a homeomorphism (linearly parameterized) that makes the target system conjugate to the reference system is learned through solving a regularized regression problem with convex barrier terms that enforces invertibility. 
Thus, bifurcation is detected when the homeomorphism can not be learned with sufficiently low error. Further research will investigate the approaches for non-local bifurcations related to limit cycles, and homoclinic or heteroclinic orbits.

\newpage
\acks{This work is supported by the faculty startup fund from NC State University. The author would also like to thank Prof. Joseph Sang-Il Kwon (Texas A\&M University) and Prof. Xunyuan Yin (Nanyang Technological University) for very beneficial discussions. During the 2022 AIChE Annual Meeting in Phoenix, Arizona, Prof. Ioannis G. Kevrekidis at Johns Hopkins University raised a question to Xunyuan on EDMD, which stimulated me in the audience to develop the idea in this paper.}
\bibliography{refs}

\begin{thebibliography}{23}
\providecommand{\natexlab}[1]{#1}
\providecommand{\url}[1]{\texttt{#1}}
\expandafter\ifx\csname urlstyle\endcsname\relax
  \providecommand{\doi}[1]{doi: #1}\else
  \providecommand{\doi}{doi: \begingroup \urlstyle{rm}\Url}\fi

\bibitem[Arnold(1988)]{arnold1988geometrical}
Vladimir~Igorevich Arnold.
\newblock \emph{Geometrical methods in the theory of ordinary differential
  equations}.
\newblock Springer, 2\textsuperscript{nd} edition, 1988.

\bibitem[Balakotaiah et~al.(1999)Balakotaiah, Dommeti, and
  Gupta]{balakotaiah1999bifurcation}
Vemuri Balakotaiah, Sandra~MS Dommeti, and Nikunj Gupta.
\newblock Bifurcation analysis of chemical reactors and reacting flows.
\newblock \emph{Chaos}, 9\penalty0 (1):\penalty0 13--35, 1999.

\bibitem[Bollt et~al.(2018)Bollt, Li, Dietrich, and
  Kevrekidis]{bollt2018matching}
Erik~M Bollt, Qianxiao Li, Felix Dietrich, and Ioannis Kevrekidis.
\newblock On matching, and even rectifying, dynamical systems through koopman
  operator eigenfunctions.
\newblock \emph{SIAM Journal on Applied Dynamical Systems}, 17\penalty0
  (2):\penalty0 1925--1960, 2018.

\bibitem[Brunton et~al.(2016)Brunton, Proctor, and
  Kutz]{brunton2016discovering}
Steven~L Brunton, Joshua~L Proctor, and J~Nathan Kutz.
\newblock Discovering governing equations from data by sparse identification of
  nonlinear dynamical systems.
\newblock \emph{Proceedings of the National Academy Sciences}, 113\penalty0
  (15):\penalty0 3932--3937, 2016.

\bibitem[Colbrook and Townsend(2024)]{colbrook2024rigorous}
Matthew~J Colbrook and Alex Townsend.
\newblock Rigorous data-driven computation of spectral properties of {Koopman}
  operators for dynamical systems.
\newblock \emph{Communications on Pure and Applied Mathematics}, 77\penalty0
  (1):\penalty0 221--283, 2024.

\bibitem[Guckenheimer and Holmes(2013)]{guckenheimer2013nonlinear}
John Guckenheimer and Philip Holmes.
\newblock \emph{Nonlinear oscillations, dynamical systems, and bifurcations of
  vector fields}.
\newblock Springer, 2013.

\bibitem[Huang and Vaidya(2022)]{huang2022convex}
Bowen Huang and Umesh Vaidya.
\newblock A convex approach to data-driven optimal control via
  {Perron-Frobenius} and {Koopman} operators.
\newblock \emph{IEEE Transactions on Automatic Control}, 67\penalty0
  (9):\penalty0 4778--4785, 2022.

\bibitem[Kaiser et~al.(2021)Kaiser, Kutz, and Brunton]{kaiser2021data}
Eurika Kaiser, J~Nathan Kutz, and Steven~L Brunton.
\newblock Data-driven discovery of {Koopman} eigenfunctions for control.
\newblock \emph{Machine Learning: Science and Technology}, 2\penalty0
  (3):\penalty0 035023, 2021.

\bibitem[Karniadakis et~al.(2021)Karniadakis, Kevrekidis, Lu, Perdikaris, Wang,
  and Yang]{karniadakis2021physics}
George~Em Karniadakis, Ioannis~G Kevrekidis, Lu~Lu, Paris Perdikaris, Sifan
  Wang, and Liu Yang.
\newblock Physics-informed machine learning.
\newblock \emph{Nature Reviews Physics}, 3\penalty0 (6):\penalty0 422--440,
  2021.

\bibitem[Kevrekidis(1987)]{kevrekidis1987numerical}
Ioannis~G Kevrekidis.
\newblock A numerical study of global bifurcations in chemical dynamics.
\newblock \emph{AIChE Journal}, 33\penalty0 (11):\penalty0 1850--1864, 1987.

\bibitem[Klus et~al.(2016)Klus, Koltai, and Sch{\"u}tte]{klus2016numerical}
Stefan Klus, Peter Koltai, and Christof Sch{\"u}tte.
\newblock On the numerical approximation of the {Perron-Frobenius} and
  {Koopman} operator.
\newblock \emph{Journal of Computational Dynamics}, 3\penalty0 (1):\penalty0
  51--77, 2016.

\bibitem[Korda and Mezi{\'c}(2018{\natexlab{a}})]{korda2018convergence}
Milan Korda and Igor Mezi{\'c}.
\newblock On convergence of extended dynamic mode decomposition to the
  {Koopman} operator.
\newblock \emph{Journal of Nonlinear Science}, 28:\penalty0 687--710,
  2018{\natexlab{a}}.

\bibitem[Korda and Mezi{\'c}(2018{\natexlab{b}})]{korda2018linear}
Milan Korda and Igor Mezi{\'c}.
\newblock Linear predictors for nonlinear dynamical systems: Koopman operator
  meets model predictive control.
\newblock \emph{Automatica}, 93:\penalty0 149--160, 2018{\natexlab{b}}.

\bibitem[Kuznetsov(2023)]{kuznetsov2023elements}
Yuri~A Kuznetsov.
\newblock \emph{Elements of applied bifurcation theory}.
\newblock Springer, 4\textsuperscript{th} edition, 2023.

\bibitem[Kuznetsov and Meijer(2019)]{kuznetsov2019numerical}
Yuri~A Kuznetsov and Hil G~E Meijer.
\newblock \emph{Numerical bifurcation analysis of maps}.
\newblock Cambridge University Press, 2019.

\bibitem[Lusch et~al.(2018)Lusch, Kutz, and Brunton]{lusch2018deep}
Bethany Lusch, J~Nathan Kutz, and Steven~L Brunton.
\newblock Deep learning for universal linear embeddings of nonlinear dynamics.
\newblock \emph{Nature Communications}, 9\penalty0 (1):\penalty0 4950, 2018.

\bibitem[Mauroy et~al.(2020)Mauroy, Susuki, and Mezi{\'c}]{mauroy2020koopman}
Alexandre Mauroy, Y~Susuki, and Igor Mezi{\'c}.
\newblock \emph{Koopman operator in systems and control}.
\newblock Springer, 2020.

\bibitem[Mezi{\'c}(2020)]{mezic2020spectrum}
Igor Mezi{\'c}.
\newblock Spectrum of the {Koopman} operator, spectral expansions in functional
  spaces, and state-space geometry.
\newblock \emph{Journal of Nonlinear Science}, 30\penalty0 (5):\penalty0
  2091--2145, 2020.

\bibitem[Narasingam et~al.(2023)Narasingam, Son, and Kwon]{narasingam2023data}
Abhinav Narasingam, Sang~Hwan Son, and Joseph Sang-Il Kwon.
\newblock Data-driven feedback stabilisation of nonlinear systems:
  {Koopman}-based model predictive control.
\newblock \emph{International Journal of Control}, 96\penalty0 (3):\penalty0
  770--781, 2023.

\bibitem[Otto and Rowley(2021)]{otto2021koopman}
Samuel~E Otto and Clarence~W Rowley.
\newblock Koopman operators for estimation and control of dynamical systems.
\newblock \emph{Annual Review on Control, Robotics, and Autonomous Systems},
  4:\penalty0 59--87, 2021.

\bibitem[Redman et~al.(2022)Redman, Fonoberova, Mohr, Kevrekidis, and
  Mezi{\'c}]{redman2022algorithmic}
William~T Redman, Maria Fonoberova, Ryan Mohr, Ioannis~G Kevrekidis, and Igor
  Mezi{\'c}.
\newblock Algorithmic (semi-) conjugacy via {Koopman} operator theory.
\newblock In \emph{2022 IEEE 61\textsuperscript{st} Conference on Decision and
  Control (CDC)}, pages 6006--6011. IEEE, 2022.

\bibitem[Strogatz(2015)]{strogatz2015nonlinear}
Steven~H Strogatz.
\newblock \emph{Nonlinear dynamics and chaos: With applications to physics,
  biology, chemistry, and engineering}.
\newblock CRC Press, 2\textsuperscript{nd} edition, 2015.

\bibitem[Williams et~al.(2015)Williams, Kevrekidis, and
  Rowley]{williams2015data}
Matthew~O Williams, Ioannis~G Kevrekidis, and Clarence~W Rowley.
\newblock A data-driven approximation of the {Koopman} operator: Extending
  dynamic mode decomposition.
\newblock \emph{Journal of Nonlinear Science}, 25\penalty0 (6):\penalty0
  1307--1346, 2015.

\end{thebibliography}

\end{document}